%% file: SADSSF.tex
\documentclass[11pt]{article}
\usepackage{amsmath}
\usepackage{amsfonts}
\usepackage{amsthm}
\usepackage{amssymb}
\usepackage{ifpdf}
\usepackage{algorithmic}
\usepackage{psfrag}
\usepackage{subfig}
\usepackage{wrapfig}
\usepackage{url}
\usepackage{nopageno}

\ifpdf

  \usepackage[pdftex]{epsfig}
  \usepackage[pdftex]{hyperref}

\else

    \usepackage[dvips]{epsfig}
    \newcommand{\href}[2]{#2}

\fi

\theoremstyle{definition}
\newtheorem{theorem}{Theorem}
\newtheorem{lemma}{Lemma}
\newtheorem*{definition}{Definition}
\newtheorem{observation}{Observation}

\newtheorem{construction}{Construction}
\newtheorem*{notation}{Notation}

\newtheorem{corollary}{Corollary}

\newcommand{\To}{\longrightarrow}

\newcommand{\Z}{\mathbb{Z}}
\newcommand{\N}{\mathbb{N}}
\newcommand{\func}[3]{#1 : #2 \rightarrow #3 }
\newcommand{\pfunc}[3]{#1 : #2 \dashrightarrow #3 }
\newcommand{\pair}[2]{(#1, #2)}

\newcommand{\color}{{\rm col}}

\newcommand{\strength}{{\rm str}}

\newcommand{\bval}[1]{[\![ #1 ]\!]}
\newcommand{\bbval}[1]{\left[\!\left[ #1 \right] \! \right]}
\newcommand{\dom}{{\rm dom} \;}
\newcommand{\asmb}{\mathcal{A}}

\newcommand{\asmbtt}{\mathcal{A}^\tau_T}
\newcommand{\ste}[2]{#1 \mapsto #2}
\newcommand{\frontier}[3]{{\partial}^{#1}_{#2}{#3}}
\newcommand{\frontiert}[1]{\partial^{\tau}{#1}}
\newcommand{\frontiertt}[1]{\frontier{\tau}{t}{#1}}

\newcommand{\frontiertau}[1]{{\partial}^{\tau}{#1}}
\newcommand{\arrowstett}[2]{#1 \xrightarrow[\tau,T]{1} #2}
\newcommand{\arrowste}[2]{#1 \stackrel{1}{\To} #2}
\newcommand{\arrowtett}[2]{#1 \xrightarrow[\tau,T]{} #2}
\newcommand{\res}[1]{\textrm{res}(#1)}
\newcommand{\termasm}[1]{\mathcal{A}_{\Box}[\mathcal{#1}]}
\newcommand{\prodasm}[1]{\mathcal{A}[\mathcal{#1}]}
\newcommand{\fgg}[1]{G^\#_{#1}}

\begin{document}

\title{\Large Self-Assembly of Discrete Self-Similar Fractals (Extended Abstract)\footnote{This research was supported in part by National
   Science Foundation Grants 0652569 and 0728806}}
\author{\small Matthew J. Patitz, and Scott M. Summers  \\ \\ \small Iowa State University, Ames, IA 50011, USA \\ \small\url{{mpatitz,summers}@cs.iastate.edu}}

\date{}

\clearpage

\maketitle
\begin{abstract}
\small In this paper, we search for {\it absolute} limitations of
the Tile Assembly Model (TAM), along with techniques to work around
such limitations. Specifically, we investigate the self-assembly of
fractal shapes in the TAM. We prove that no self-similar fractal
fully weakly self-assembles at temperature 1, and that certain kinds
of self-similar fractals do not strictly self-assemble at any
temperature. Additionally, we extend the fiber construction from
Lathrop et. al. (2007) to show that any self-similar fractal
belonging to a particular class of ``nice'' self-similar fractals
has a fibered version that strictly self-assembles in the TAM.
\end{abstract}

\input{section1}
\input{section2}
\input{section3}
\input{section4}
\input{section5}
\input{section6}

\subsubsection*{Acknowledgment}
We thank Dave Doty, Jim Lathrop, Jack Lutz, and Aaron Sterling for
useful discussions.

\small
\bibliographystyle{amsplain}
\bibliography{main,dim,random,dimrelated,rbm}

\end{document}

%% file: section1.tex
\section{Introduction}
Self-assembly is a bottom-up process by which (usually a small
number of) fundamental components automatically coalesce to form a
target structure. In 1998, Winfree \cite{Winf98} introduced the
(abstract) Tile Assembly Model (TAM) - an extension of Wang tiling
\cite{Wang61,Wang63}, and a mathematical model of the DNA
self-assembly pioneered by Seeman et. al. \cite{Seem82}. In the TAM,
the fundamental components are un-rotatable, but translatable ``tile
types'' whose sides are labeled with glue ``colors'' and
``strengths.'' Two tiles that are placed next to each other {\it
interact} if the glue colors on their abutting sides match, and they
{\it bind} if the strength on their abutting sides matches, and is
at least a certain ``temperature.'' Rothemund and Winfree
\cite{RotWin00,Roth01} later refined the model, and Lathrop et. al.
\cite{SSADST} gave a treatment of the TAM in which equal status is
bestowed upon the self-assembly of infinite and finite structures.
There are also several generalizations \cite{AGKS04,MLR07,KS07} of
the TAM.

Despite its deliberate over-simplification, the TAM is a
computationally and geometrically expressive model. For instance,
Winfree \cite{Winf98} proved that the TAM is computationally
universal, and thus can be directed algorithmically. Winfree
\cite{Winf98} also exhibited a seven-tile-type self-assembly system,
directed by a clever XOR-like algorithm, that ``paints'' a picture
of a well-known shape, the discrete Sierpinski triangle
$\mathbf{S}$, onto the first quadrant. Note that the underlying {\it
shapes} of each of the previous results are actually infinite
canvases that completely cover the first quadrant, onto which
computationally interesting shapes are painted (i.e., full weak
self-assembly). Moreover, Lathrop et. al \cite{CCSA} recently gave a
new characterization of the computably enumerable sets in terms of
weak self-assembly using a ``ray construction.'' It is natural to
ask the question: How expressive is the TAM with respect to the
self-assembly of a particular, possibly infinite shape, and nothing
else (i.e., strict self-assembly)?

In the case of strict self-assembly of finite shapes, the TAM
certainly remains an interesting model, so long as the size (tile
complexity) of the assembly system is required to be ``small''
relative to the shape that it ultimately produces. For instance,
Rothemund and Winfree \cite{RotWin00} proved that there are small
tile sets in which large squares self-assemble. Moreover,
Soloveichik and Winfree \cite{SolWin07} established the remarkable
fact that, if one is not concerned with the scale of an
``algorithmically describable'' finite shape, then there is always a
small tile set in which the shape self-assembles. Note that if the
tile complexity of an assembly system is unbounded, then every
finite shape trivially (but perhaps not feasibly) self-assembles.

When the tile complexity of an assembly system is unbounded (yet
finite), only infinite shapes are of interest. In the case of strict
self-assembly of infinite shapes, the power of the TAM has only
recently been investigated. Lathrop et. al. \cite{SSADST}
established that self-similar tree shapes do not strictly
self-assemble in the TAM given any finite number of tile types. A
``fiber construction''  is also given in \cite{SSADST}, which
strictly self-assembles a non-trivial fractal structure.

In this paper, we search for (1) {\it absolute} limitations of the
TAM, with respect to the strict self-assembly of shapes, and (2)
techniques that allow one to ``work around'' such limitations.
Specifically, we investigate the strict self-assembly of fractal
shapes in the TAM. We prove three main results: two negative and one
positive. Our first negative (i.e., impossibility) result says that
no self-similar fractal fully weakly self-assembles in the TAM (at
temperature 1). In our second impossibility result, we exhibit a
class of discrete self-similar fractals, to which the standard
discrete Sierpinski triangle belongs, that do not strictly
self-assemble in the TAM (at {\it any} temperature). Finally, in our
positive result, we use simple modified counters to extend the fiber
construction from Lathrop et. al. \cite{SSADST} to a particular
class of discrete self-similar fractals.

%% file: section2.tex
\section{Preliminaries}
\subsection{Notation and Terminology}
We work in the discrete Euclidean plane $\Z^2 = \Z \times \Z$. We
write $U_2$ for the set of all {\it unit vectors}, i.e., vectors of
length $1$, in $\Z^2$.  We regard the four elements of $U_2$ as
(names of the cardinal) {\it directions} in $\Z^2$.

We write $[X]^2$ for the set of all $2$-element subsets of a set
$X$.  All {\it graphs} here are undirected graphs, i.e., ordered
pairs $G = (V, E)$, where $V$ is the set of {\it vertices} and $E
\subseteq [V]^2$ is the set of {\it edges}. A {\it cut} of a graph
$G=(V,E)$ is a partition $C=(C_0,C_1)$ of $V$ into two nonempty,
disjoint subsets $C_0$ and $C_1$.

A {\it binding function} on a graph $G = (V,E)$ is a function
$\beta:E\rightarrow \mathbb{N}$. (Intuitively, if $\{u,v\} \in E$,
then $\beta\left(\{u,v\}\right)$ is the strength with which $u$ is
bound to $v$ by $\{u,v\}$ according to $\beta$. If $\beta$ is a
binding function on a graph $G=(V,E)$ and $C=(C_0,C_1)$ is a cut of
$G$, then the {\it binding strength} of $\beta$ on $C$ is
$$
\beta_C = \left\{ \beta(e) \left| e\in E, e\cap C_0 \ne
\emptyset,\textmd{ and } e \cap C_1\ne \emptyset \right.\right\}.
$$
The {\it binding strength} of $\beta$ on the graph $G$ is then
$$
\beta(G) = \min\left\{ \beta_C \left| C \textmd{ is a cut of } G
\right. \right\}.
$$

A {\it binding graph} is an ordered triple $G=(V,E,\beta)$, where
$(V,E)$ is a graph and $\beta$ is a binding function on $(V,E)$. If
$\tau \in \mathbb{N}$, then a binding graph $G = (V,E,\beta)$ is
$\tau$-{\it stable} if $\beta(V,E)\geq \tau$.

A {\it grid graph} is a graph $G = (V, E)$ in which $V \subseteq
\Z^2$ and every edge $\{\vec{m}, \vec{n} \} \in E$ has the property
that $\vec{m} - \vec{n} \in U_2$.  The {\it full grid graph} on a
set $V \subseteq \Z^2$ is the graph $\fgg{V} = (V, E)$ in which $E$
contains {\it every} $\{\vec{m}, \vec{n} \} \in [V]^2$ such that
$\vec{m} - \vec{n} \in U_2$.

We say that $f$ is a {\it partial function} from a set $X$ to a set
$Y$, and we write $\pfunc{f}{X}{Y}$, if $f: D\rightarrow Y$ for some
set $D \subseteq X$. In this case, $D$ is  the {\it domain} of $f$,
and we write $D = \dom{f}$.

All logarithms here are base-2.

\subsection{The Tile Assembly Model}
We review the basic ideas of the Tile Assembly Model. Our
development largely follows that of \cite{RotWin00,Roth01}, but some
of our terminology and notation are specifically tailored to our
objectives.  In particular, our version of the model only uses
nonnegative ``glue strengths'', and it bestows equal status on
finite and infinite assemblies. We emphasize that the results in
this section have been known for years, e.g., they appear, with
proofs, in \cite{Roth01}.

\begin{definition}
A {\it tile type} over an alphabet $\Sigma$ is a function
$\func{t}{U_2}{\Sigma^* \times \N}$. We write
$t=\pair{\color_t}{\strength_t}$, where
$\func{\color_t}{U_2}{\Sigma^*}$, and $\func{\strength_t}{U_2}{\N}$
are defined by $t(\vec{u}) =
\pair{\color_t(\vec{u})}{\strength_t(\vec{u})}$ for all $\vec{u} \in
U_2$.
\end{definition}

Intuitively, a tile of type $t$ is a unit square. It can be
translated but not rotated, so it has a well-defined ``side
$\vec{u}\;$'' for each $\vec{u} \in U_2$.  Each side $\vec{u}$ of
the tile is covered with a ``glue'' of {\it color}
$\color_t(\vec{u})$ and {\it strength} $\strength_t(\vec{u})$.  If
tiles of types $t$ and $t^\prime$ are placed with their centers at
$\vec{m}$ and $\vec{m}+\vec{u}$, respectively, where $\vec{m} \in
\Z^2$ and $\vec{u} \in U_2$, then they will {\it bind} with strength
$\strength_t(\vec{u}) \cdot \bval{t(\vec{u}) = t^\prime(-\vec{u})}$
where $\bval{\phi}$ is the {\it Boolean} value of the statement
$\phi$. Note that this binding strength is $0$ unless the adjoining
sides have glues of both the same color and the same strength.

For the remainder of this section, unless otherwise specified, $T$
is an arbitrary set of tile types, and $\tau \in \N$ is the
``temperature.''

\begin{definition}
A T-{\it configuration} is a partial function
$\pfunc{\alpha}{\Z^2}{T}$.
\end{definition}

Intuitively, a configuration is an assignment $\alpha$ in which a
tile of type $\alpha(\vec{m})$ has been placed (with its center) at
each point $\vec{m} \in \dom \alpha$. The following data structure
characterizes how these tiles are bound to one another.

\begin{definition}
The {\it binding graph of} a $T$-configuration
$\pfunc{\alpha}{\Z^2}{T}$ is the binding graph $G_\alpha = (V, E,
\beta )$, where $(V, E)$ is the grid graph given by
\begin{alignat*}{1}
V = \; &\dom \alpha \text{,} \\
E = \; &\left\{\left. \{\vec{m}, \vec{n}\} \in {[V]}^2 \right|
\vec{m} - \vec{n} \in U_n,  \color_{\alpha(\vec{m})}\left(\vec{n} -
\vec{m}\right) = \color_{\alpha(\vec{n})}\left(\vec{m} -
\vec{n}\right), \text{and } \strength_{\alpha(\vec{m})}\left(\vec{n}
-\vec{m}\right) > 0 \right\},
\end{alignat*}
and the binding function $\func{\beta}{E}{\Z^+}$ is given by
\[
\beta\left(\{\vec{m}, \vec{n}\}\right) =
\strength_{\alpha(\vec{m})}\left(\vec{n} -\vec{m}\right)
\]
for all $\left\{\vec{m}, \vec{n} \right\} \in E$.
\end{definition}

\begin{definition} \textmd{ }
\begin{enumerate}
  \item [1.]  A $T$-configuration $\alpha$ is $\tau$-{\it stable} if its binding
  graph $G_\alpha$ is $\tau$-stable.
  \item [2.]  A $\tau$-$T$-{\it assembly} is a $T$-configuration that is
  $\tau$-stable.  We write $\asmb^\tau_T$ for the set of all $\tau$-$T$-assemblies.
\end{enumerate}
\end{definition}

\begin{definition}
Let $\alpha$ and $\alpha^\prime$ be $T$-configurations.
\begin{enumerate}
  \item [1.]  $\alpha$ is a {\it subconfiguration} of $\alpha^\prime$, and we
  write $\alpha \sqsubseteq \alpha^\prime$, if $\dom \alpha \subseteq \dom
  \alpha^\prime$ and, for all $\vec{m} \in \dom \alpha$, $\alpha(\vec{m}) =
  \alpha^\prime(\vec{m}).$
  \item [2.]  $\alpha^\prime$ is a {\it single-tile extension} of $\alpha$ if
  $\alpha \sqsubseteq \alpha^\prime$ and $\dom \alpha^\prime - \dom \alpha$ is a
  singleton set.  In this case, we write $\alpha^\prime = \alpha +
  (\ste{\vec{m}}{t})$, where $\{\vec{m}\} = \dom \alpha^\prime - \dom \alpha$
  and $t = \alpha^\prime(\vec{m})$.
\end{enumerate}
\end{definition}

Note that the expression $\alpha + (\ste{\vec{m}}{t})$ is only
defined when $\vec{m} \in \Z^2 - \dom \alpha$.

We next define the ``$\tau$-$t$-frontier'' of a $\tau$-$T$-assembly
$\alpha$ to be the set of all positions at which a tile of type $t$
can be ``$\tau$-stably added'' to the assembly $\alpha$.

\begin{definition}
Let $\alpha \in \asmb^\tau_T$.
\begin{enumerate}
  \item[1.]  For each $t \in T$, the $\tau$-$t$-{\it frontier} of $\alpha$ is
  the set
\[
\frontiertt{\alpha} = \left\{ \vec{m} \in \Z^2-\dom \alpha \left| \;
\sum_{\vec{u} \in U_2} \strength_{t}(\vec{u}) \cdot
\bbval{\alpha(\vec{m} + \vec{u})(-\vec{u}) = t(\vec{u})} \geq \tau
\right. \right\}.
\]
  \item[2.] The $\tau$-{\it frontier} of $\alpha$ is the set
\begin{equation*}
\frontiertau{\alpha} = \bigcup_{t \in T} \frontiertt{\alpha}.
\end{equation*}
\end{enumerate}
\end{definition}

The following lemma shows that the definition of
$\frontiertt{\alpha}$ achieves the desired effect.

\begin{lemma}
Let $\alpha \in \asmbtt$, $\vec{m} \in \mathbb{Z}^2 - \dom \alpha$,
and $t \in T$. Then $\alpha + (\ste{\vec{m}}{t}) \in \asmb^\tau_T$
if and only if $\vec{m} \in \frontiertt{\alpha}$.
\end{lemma}

\begin{notation}
We write $\arrowstett{\alpha}{\alpha^\prime}$ (or, when $\tau$ and
$T$ are clear from context, $\arrowste{\alpha}{\alpha^\prime}$) to
indicate that $\alpha, \alpha^\prime \in \asmbtt$ and
$\alpha^\prime$ is a single-tile extension of $\alpha$.
\end{notation}

In general, self-assembly occurs with tiles adsorbing
nondeterministically and asynchronously to a growing assembly.  We
now define assembly sequences, which are particular ``execution
traces'' of how this might occur.

\begin{definition}
A $\tau$-$T$-{\it assembly sequence} is a sequence
$\vec{\alpha}=(\alpha_i  \mid  0 \leq i<k)$ in $\asmbtt$, where $k
\in \mathbb{Z}^+ \cup \{\infty\}$ and, for each $i$ with $1 \leq i+1
< k$, $\arrowstett{\alpha_i}{\alpha_{i+1}}$.
\end{definition}

Note that assembly sequences may be finite or infinite in length.
Note also that, in any $\tau$-$T$-assembly sequence
$\vec{\alpha}=(\alpha_i  \mid  0 \leq i < k)$, we have $\alpha_i
\sqsubseteq \alpha_j$ for all $0 \leq i \leq j < k$.

\begin{definition}
The {\it result} of a $\tau$-$T$-assembly sequence
$\vec{\alpha}=(\alpha_i  \mid  0 \leq i < k)$ is the unique
$T$-configuration $\alpha=\res{\vec{\alpha}}$ satisfying $\dom
\alpha = \bigcup_{0 \leq i < k}{\dom \alpha_i}$ and $\alpha_i
\sqsubseteq \alpha$ for each $0 \leq i < k$.
\end{definition}

It is clear that $\res{\vec{\alpha}} \in \asmbtt$ for every
$\tau$-$T$-assembly sequence $\vec{\alpha}$.

\begin{definition}
Let $\alpha, \alpha^\prime \in \asmbtt$.
\begin{enumerate}
\item[1.] A $\tau$-$T$-{\it assembly sequence from} $\alpha$ {\it
to} $\alpha^\prime$ is a $\tau$-$T$-assembly sequence
$\vec{\alpha}=(\alpha_i  \mid  0 \leq i < k)$ such that $\alpha_0 =
\alpha$ and $\res{\vec{\alpha}} = \alpha^\prime$.
\item[2.] We write $\arrowtett{\alpha}{\alpha^\prime}$ (or, when
$\tau$ and $T$ are clear from context, $\alpha \To \alpha^\prime$)
to indicate that there exists a $\tau$-$T$-assembly sequence from
$\alpha$ to $\alpha^\prime$.
\end{enumerate}
\end{definition}



\begin{definition}
An assembly $\alpha \in \asmbtt$ is {\it terminal} if and only if
$\frontiert{\alpha} = \emptyset$.
\end{definition}




We now define tile assembly systems.

\begin{definition} \textmd{ }
\begin{enumerate}
\item[1.] A {\it generalized tile assembly system}
({\it GTAS}) is an ordered triple
$$
\mathcal{T} = (T,\sigma,\tau),
$$
where $T$ is a set of tile types, $\sigma \in \asmbtt$ is the {\it
seed assembly}, and $\tau \in \mathbb{N}$ is the {\it temperature}.
\item[2.] A {\it tile assembly system} ({\it TAS}) is a GTAS $\mathcal{T} = (T,\sigma,\tau)$
in which the sets $T$ and $\dom \sigma$ are finite.
\end{enumerate}
\end{definition}

Intuitively, a ``run'' of a GTAS $\mathcal{T}=(T,\sigma,\tau)$ is
any $\tau$-$T$-assembly sequence $\vec{\alpha} = (\alpha_i \mid 0
\leq i < k)$ that begins with $\alpha_0 = \sigma$. Accordingly, we
define the following sets.

\begin{definition} Let $\mathcal{T} = (T,\sigma,\tau)$ be a GTAS.
\begin{enumerate}
\item[1.] The {\it set of assemblies produced by} $\mathcal{T}$ is
$$
\prodasm{T} = \left\{ \alpha \in \asmbtt \left|
\arrowtett{\sigma}{\alpha}
 \right. \right\}.
$$
\item[2.] The {\it set of terminal assemblies produced by}
$\mathcal{T}$ is
$$
\termasm{T} = \left\{ \left. \alpha \in \mathcal{A}[\mathcal{T}]
\right| \alpha\textrm{ is terminal} \right\}.
$$
\end{enumerate}
\end{definition}

\begin{definition}
A GTAS $\mathcal{T} = (T,\sigma,\tau)$ is {\it directed} if $\left|
\termasm{T} \right|=1$.
\end{definition}

We are using the terminology of the mathematical theory of relations
here.  The reader is cautioned that the term "directed" has also
been used for a different, more specialized notion in self-assembly
\cite{AKKR02}.



In the present paper, we are primarily interested in the
self-assembly of sets.

\begin{definition}
Let $\mathcal{T} = (T,\sigma,\tau)$ be a GTAS, and let $X \subseteq
\mathbb{Z}^2$.
\begin{enumerate}
\item The set $X$ {\it weakly self-assembles} in $\mathcal{T}$
if there is a set $B \subseteq T$ such that, for all $\alpha \in
\termasm{T}$, $\alpha^{-1}(B) = X$.

\item The set $X$ {\it fully weakly self-assembles} in $\mathcal{T}$
if $X$ and $\mathbb{Z}^2-X$ both weakly self-assemble.

\item The set $X$ {\it strictly self-assembles} in $\mathcal{T}$
if, for all $\alpha \in \termasm{T}$, $\dom \alpha = X$.
\end{enumerate}
\end{definition}

Intuitively, a set $X$ weakly self-assembles in $\mathcal{T}$ if
there is a designated set $B$ of ``black'' tile types such that
every terminal assembly of $\mathcal{T}$ ``paints the set $X$ - and
only the set $X$ - black''. Moreover, a set fully weakly
self-assembles in $\mathcal{T}$ if every terminal assembly of
$\mathcal{T}$ tiles the {\it entire} plane on top of which $X$ is
``painted.'' In contrast, a set $X$ strictly self-assembles in
$\mathcal{T}$ if every terminal assembly of $\mathcal{T}$ has tiles
on the set $X$ and only on the set $X$. Clearly, every set that
strictly self-assembles in a GTAS $\mathcal{T}$ also weakly
self-assembles in $\mathcal{T}$.

We now have the machinery to say what it means for a set in the
discrete Euclidean plane to self-assemble in either the fully weak,
weak or the strict sense.

\begin{definition} Let $X \subseteq \mathbb{Z}^2$.
\begin{enumerate}
\item The set $X$ {\it weakly self-assembles} if there is a TAS $\mathcal{T}$ such that $X$ weakly self-assembles in
$\mathcal{T}$.

\item The set $X$ {\it fully weakly self-assembles} if there is a
TAS $\mathcal{T}$ such that $X$ fully weakly self-assembles in
$\mathcal{T}$.

\item The set $X$ {\it strictly self-assembles} if there is a TAS $\mathcal{T}$ such that $X$ strictly self-assembles in
$\mathcal{T}$.
\end{enumerate}
\end{definition}

Note that $\mathcal{T}$ is required to be a TAS, i.e., finite, in
all three parts of the above definition.

\subsection{Local Determinism}
The proof of our construction uses the local determinism method of
Soloveichik and Winfree \cite{SolWin07}, which we now review.

\begin{notation}
For each $T$-configuration $\alpha$, each $\vec{m} \in
\mathbb{Z}^2$, and each $\vec{u} \in U_2$,
$$
\text{str}_{\alpha}(\vec{m},\vec{u}) =
\text{str}_{\alpha(\vec{m})}(\vec{u})\cdot
\bval{\alpha(\vec{m})(\vec{u}) = \alpha(\vec{m}+\vec{u})(-\vec{u})}.
$$
(The Boolean value on the right is 0 if $\{\vec{m},\vec{m}+\vec{u}\}
\nsubseteq \dom{\alpha}$.)
\end{notation}

\begin{notation}
If $\vec{\alpha} = (\alpha_i \; | \; 0\leq i<k)$ is a
$\tau$-$T$-assembly sequence and $\vec{m} \in \mathbb{Z}^2$, then
the $\vec{\alpha}$-{\it index} of $\vec{m}$ is
$$
i_{\vec{\alpha}}(\vec{m}) = \min\{ i\in \mathbb{N} \; \left| \;
\vec{m} \in \dom{\alpha_i} \right. \}.
$$
\end{notation}

\begin{observation} $\vec{m} \in \dom{\res{\vec{\alpha}}} \Leftrightarrow
i_{\vec{\alpha}}(\vec{m}) < \infty$.
\end{observation}

\begin{notation}
If $\vec{\alpha} = (\alpha_i \; | \; 0\leq i<k)$ is a
$\tau$-$T$-assembly sequence, then, for $\vec{m},\vec{m}' \in
\mathbb{Z}^2$,
$$
\vec{m} \prec_{\vec{\alpha}} \vec{m}' \Leftrightarrow
i_{\vec{\alpha}}(\vec{m}) < i_{\vec{\alpha}}(\vec{m}').
$$
\end{notation}

\begin{definition}
\label{local_determinism_sets} (Soloveichik and Winfree
\cite{SolWin07}) Let $\vec{\alpha} = (\alpha_i \; | \; 0\leq i < k)$
be a $\tau$-$T$-assembly sequence, and let $\alpha =
\res{\vec{\alpha}}$. For each location $\vec{m} \in \dom{\alpha}$,
define the following sets of directions.
\begin{enumerate}
\item[1.] $\textmd{IN}^{\vec{\alpha}}(\vec{m}) = \left\{ \vec{u} \in U_2 \left| \vec{m}+\vec{u} \prec_{\vec{\alpha}} \vec{m} \textmd{ and }
\textmd{str}_{\alpha_{i_{\vec{\alpha}}(\vec{m})}}(\vec{m},\vec{u})>0
\right.\right\}$.
\item[2.] $\textmd{OUT}^{\vec{\alpha}}(\vec{m}) = \left\{ \vec{u}\in U_2 \left|
 -\vec{u} \in \textmd{IN}^{\vec{\alpha}}(\vec{m}+\vec{u} \right.)
\right\}$.
\end{enumerate}
\end{definition}

Intuitively, $\textmd{IN}^{\vec{\alpha}}(\vec{m})$ is the set of
sides on which the tile at $\vec{m}$ initially binds in the assembly
sequence $\vec{\alpha}$, and $\textmd{OUT}^{\vec{\alpha}}(\vec{m})$
is the set of sides on which this tile propagates information to
future tiles.

Note that $\textmd{IN}^{\vec{\alpha}}(\vec{m}) = \emptyset$ for all
$\vec{m} \in \alpha_0$.

\begin{notation}
If $\vec{\alpha} = (\alpha_i \; | \; 0 \leq i < k)$ is a
$\tau$-$T$-assembly sequence, $\alpha = \res{\vec{\alpha}}$, and
$\vec{m} \in \dom{\alpha} - \dom{\alpha_0}$, then
$$
\vec{\alpha} \setminus \vec{m} = \alpha \upharpoonright
\left(\dom{\alpha} - \{\vec{m}\} -
\left(\vec{m}+\textmd{OUT}^{\vec{\alpha}}(\vec{m})\right)\right).
$$
\end{notation}

(Note that $\vec{\alpha} \setminus \vec{m}$ is a $T$-configuration
that may or may not be a $\tau$-$T$-assembly.

\begin{definition}
\label{local_determinism_definition} (Soloveichik and Winfree
\cite{SolWin07}). A $\tau$-$T$-assembly sequence $\vec{\alpha} =
(\alpha_i \; | \; 0 \leq i < k)$ with result $\alpha$ is {\it
locally deterministic} if it has the following three properties.
\begin{enumerate}
\item[1.] For all $\vec{m} \in \dom{\alpha} - \dom{\alpha_0}$,
$$
\sum_{\vec{u} \in
\textmd{IN}^{\vec{\alpha}}(\vec{m})}{\textmd{str}_{\alpha_{i_{\vec{\alpha}}(\vec{m})}}(\vec{m},\vec{u})
} = \tau.
$$
\item[2.] For all $\vec{m} \in \dom{\alpha} - \dom{\alpha_0}$ and
all $t \in T-\{\alpha(\vec{m})\}$, $\vec{m} \not \in
\frontiertt{\left(\vec{\alpha} \setminus \vec{m}\right)}$.
\item[3.] $\frontiert{\alpha} = \emptyset$.
\end{enumerate}
\end{definition}

That is, $\vec{\alpha}$ is locally deterministic if (1) each tile
added in $\vec{\alpha}$ ``just barely'' binds to the assembly; (2)
if a tile of type $t_0$ at a location $\vec{m}$ and its immediate
``OUT-neighbors'' are deleted from the {\it result} of
$\vec{\alpha}$, then no tile of type $t \ne t_0$ can attach itself
to the thus-obtained configuration at location $\vec{m}$; and (3)
the result of $\vec{\alpha}$ is terminal.

\begin{definition}
\label{locally_deterministic_tas_def} A GTAS $\mathcal{T} =
(T,\sigma,\tau)$ is {\it locally deterministic} if there exists a
locally determinstic $\tau$-$T$-assembly sequence
$\vec{\alpha}=(\alpha_i \; | \; 0\leq i < k)$ with $\alpha_0 =
\sigma$.
\end{definition}

\begin{theorem}
\label{local_determinism_theorem} (Soloveichik and Winfree
\cite{SolWin07}) Every locally deterministic GTAS is directed.
\end{theorem}

\subsection{Discrete Self-Similar Fractals}
In this subsection we introduce discrete self-similar fractals.

\begin{definition}
Let $1 < c \in \mathbb{N}$, and $X\subsetneq \mathbb{N}^2$ (we do not
consider $\mathbb{N}^2$ to be a self-similar fractal). We say that
$X$ is a $c$-{\it discrete self-similar fractal}, if there is a set
$\{(i,i) \mid i \in \{0,\ldots,c-1\}\} \ne V \subseteq
\{0,\ldots,c-1\}\times\{0,\ldots,c-1\}$ such that
$$
X = \bigcup_{i=0}^{\infty}{X_i},
$$
where $X_i$ is the $i^{\textmd{th}}$ {\it stage} satisfying $X_0 =
\{(0,0)\}$, and $X_{i+1} = X_i \cup \left(X_i + c^i V \right)$. In
this case, we say that $V$ {\it generates} $X$. $X$ is a {\it
discrete self-similar fractal} if it is a $c$-discrete self-similar
fractal for some $c \in \mathbb{N}$.
\end{definition}

In this paper, we are concerned with the following class of
self-similar fractals.

\begin{definition}
A {\it nice discrete self-similar fractal} is a discrete
self-similar fractal such that $(\{0,\ldots,c-1\} \times \{0\}) \cup
(\{0\}\times\{0,\ldots,c-1\}) \subseteq V$, and $\fgg{V}$ is
connected.
\end{definition}
\begin{figure}[htp]
  \begin{center}
  \subfloat[Nice]{\label{fig:nice}\includegraphics[width=1.25in]{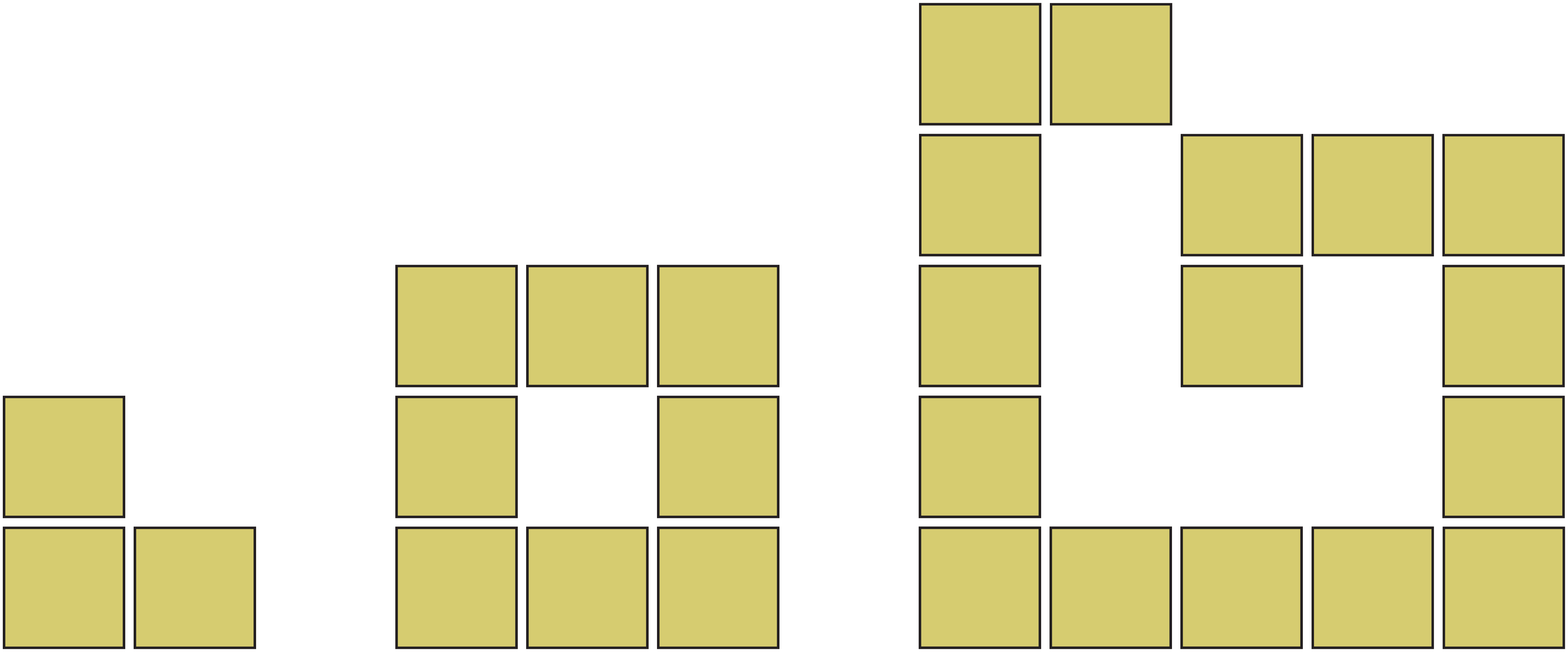}}
  \quad\quad
  \subfloat[Non-nice]{\label{fig:bad}\includegraphics[width=0.73in]{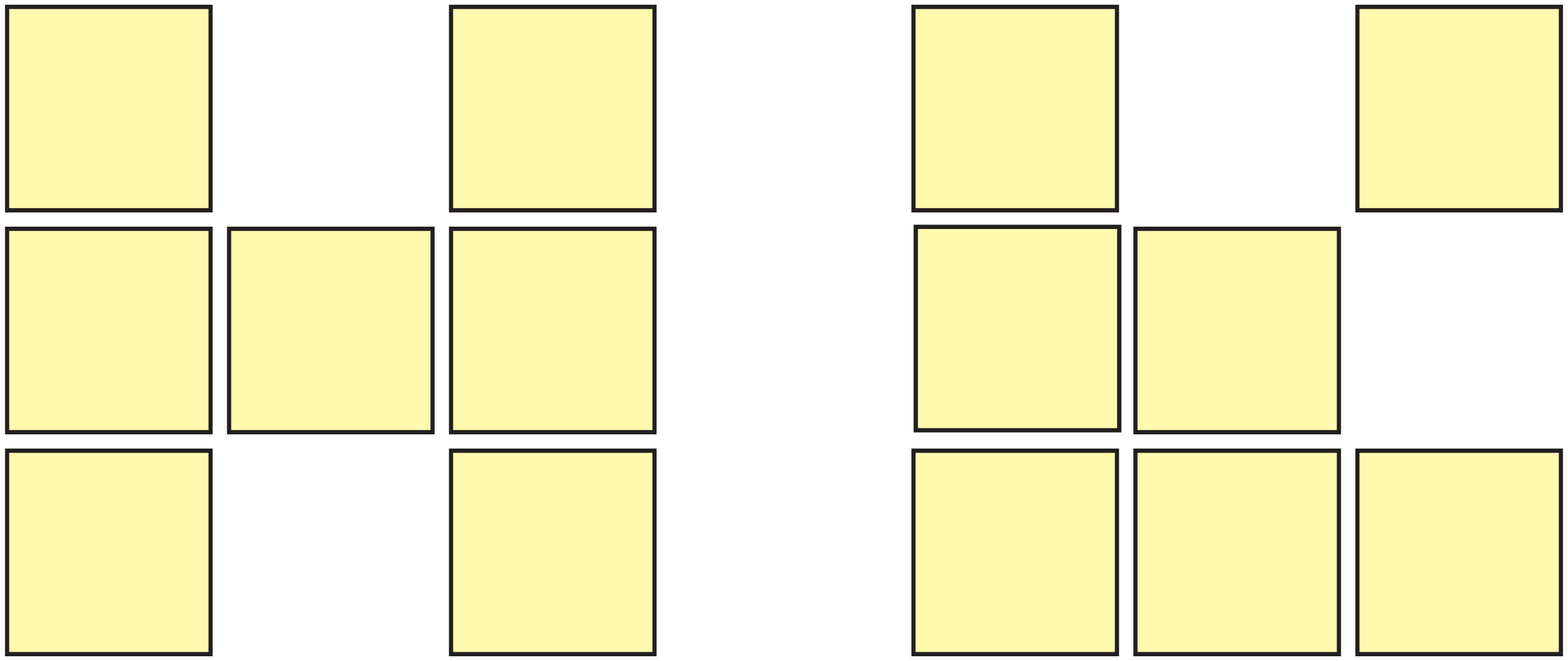}}
  \caption{\small The first stages of discrete self-similar fractals. The fractals in (a) are nice, whereas (b) shows two non-nice fractals.}
  \label{fig:fractals}
  \vspace{-20pt}
  \end{center}
\end{figure}

\subsection{Zeta-Dimension}
The most commonly used dimension for discrete fractals is
zeta-dimension, which we use in this paper. The discrete-continuous
correspondence mentioned in the introduction preserves dimension
somewhat generally. Thus, for example, the zeta-dimension of the
discrete Sierpinski triangle is the same as the Hausdorff dimension
of the continuous Sierpinski triangle.

Zeta-dimension has been re-discovered several times by researchers
in various fields over the past few decades, but its origins
actually lie in Euler's (real-valued predecessor of the Riemann)
zeta-function \cite{Euler1737} and Dirichlet series.  For each set
$A \subseteq \Z^2$, define the {\it A-zeta-function} $\zeta_A:[0,
\infty)\rightarrow[0, \infty]$ by $\zeta_A(s) = \sum_{(0, 0) \ne (m,
n) \in A} (|m|+|n|)^{-s}$ for all $s \in [0,\infty)$. Then the {\it
zeta-dimension} of $A$ is
\[
     \textmd{Dim}_\zeta(A) = \inf \{s \; | \; \zeta_A(s) < \infty\}.
\]
It is clear that $0 \le Dim_\zeta(A) \le 2$ for all $A \subseteq
\Z^2$. It is also easy to see (and was proven by Cahen in 1894; see
also \cite{Apos97,HarWri79}) that zeta-dimension admits the
``entropy characterization''
\begin{gather*}
\tag*{(2.1)}
     \textmd{Dim}_\zeta(A) = \limsup_{n \rightarrow \infty}\frac{ \log|A_{\le n}|}{\log
     n},
\end{gather*}
where $A_{\le n} = \{(k,l) \in A \mid  |k|+|l| \le n\}$. Various
properties of zeta-dimension, along with extensive historical
citations, appear in the recent paper \cite{ZD}, but our technical
arguments here can be followed without reference to this material.
We use the fact, verifiable by routine calculation, that (2.1) can
be transformed by changes of variable up to exponential, e.g.,
\[
{\rm Dim}_\zeta(A) = \limsup_{n \rightarrow \infty} \frac{ \log
|A_{[0,2^n] \cap \mathbb{N}}|}{n}
\]
also holds.

%% file: section3.tex
\section{Impossibility Results}
In this section, we explore the theoretical limitations of the Tile
Assembly Model with respect to the self-assembly of fractal shapes.
First, we establish that no discrete self-similar fractal fully
weakly self-assembles at temperature $\tau = 1$. Second, we exhibit
a class $\mathcal{C}$ of discrete self-similar fractals, and prove
that if $F \in \mathcal{C}$, then $F$ does not strictly
self-assemble in the TAM.

\begin{definition}(Lathrop et. al. \cite{SSADST})
Let $G = (V,E)$ be a graph, and let $D \subseteq V$.
For each $r \in V$, the $D$-$r$-{\it rooted subgraph} of
$G$ is the graph $G_{D,r} = \left( V_{D,r}, E_{D,r} \right)$, where
$$
V_{D,r} = \left\{ v \in V \left| \textmd{ every simple path from } v
\textmd{ to (any vertex in) } D \textmd{ in } G \textmd{ goes
through } r \right. \right\}
$$
and $E_{D,r} = E \cap \left[ V_{D,r} \right]^2$. $B$ is a $D$-{\it
subgraph} of $G$ if it is a $D$-$r$-rooted subgraph of $G$ for some
$r \in V$.

\end{definition}

\begin{definition} Let $G = (V,E)$ be a graph. Fix a
set $D \subseteq V$, and let $r,r' \in V$.
\begin{enumerate}
\item (Adleman et. al. \cite{ACGHKMR02})  $G_{D,r}$ is {\it
isomorphic} to $G_{D,r'}$, and we write $G_{D,r} \sim G_{D,r'}$ if
there exists a vector $\vec{a} \in \mathbb{Z}^2$ such that $V_{D,r}
= V_{D,r'} + \vec{a}$.

\item We say that $G_{D,r}$ is {\it unique} if, for all $r' \in V$, $G_{D,r} \sim
G_{D,r'} \Rightarrow r = r'$.
\end{enumerate}
\end{definition}

We will use the following technical result to prove that no
self-similar fractal weakly self-assembles at temperature $\tau=1$.

\begin{lemma}\label{pathlemma}(Adleman et. al. \cite{ACGHKMR02}) Let $X \subsetneq
\mathbb{N}^2$ such that $\fgg{X}$ is a finite tree, and assume that
$X$ strictly self-assembles in the TAS $\mathcal{T} =
(T,\sigma,\tau)$. Let $\alpha \in \termasm{\mathcal{T}}$. If
$\alpha\left(\vec{u}\right) = \alpha\left(\vec{v}\right)$, then the
$G_{\dom{\sigma},\vec{u}} \sim G_{\dom{\sigma},\vec{v}}$.
\end{lemma}

The following construction says that if it is possible to
self-assemble a finite path $P$ at temperature 1 (not necessarily
uniquely), then there is always a TAS $\mathcal{T}_P$ in which $P$
uniquely self-assembles at temperature 1.

\begin{construction} \label{crapfactory2} Let $T$ be a finite set of tile types, and
$\vec{\alpha} = (\alpha_i \mid 0 \leq i < k)$ be a $1$-$T$-assembly
sequence, with $\alpha = \res{\vec{\alpha}}$, satisfying
\begin{enumerate}
\item $\dom{\alpha_0} = \{(0,0)\}$, and
\item $\fgg{\alpha}$ is a connected, finite path $P$.
\end{enumerate}
It is clear that for all $\vec{v} \in P$,
$\left|\textmd{IN}^{\vec{\alpha}}(\vec{v}) \right| = \left|
\textmd{OUT}^{\vec{\alpha}}(\vec{v})\right| = 1$. Now define, for
each $\vec{v} \in P$, the (unique) vectors
$\vec{v}_{\textmd{in}},\vec{v}_{\textmd{out}}$, satisfying
$\vec{v}_{\textmd{in}} \in \textmd{IN}^{\vec{\alpha}}(\vec{v})$, and
$\vec{v}_{\textmd{out}} \in \textmd{OUT}^{\vec{\alpha}}(\vec{v})$.
For each $\vec{v} \in P$, define the tile type $t_{\vec{v}}$, where
for all $\vec{u} \in U_2$,
$$
t_{\vec{v}}(\vec{u}) = \left\{
\begin{array}{ll}
\left(\color_{\alpha(\vec{v})}(\vec{u})\cdot\textmd{``in''},\strength_{\alpha(\vec{v})}(\vec{u})\right) & \textrm{ if } \vec{v}+\vec{u} = \vec{v}_{\textmd{in}} \\
\left(\color_{\alpha(\vec{v})}(\vec{u})\cdot\textmd{``out''},\strength_{\alpha(\vec{v})}(\vec{u})\right) & \textrm{ if } \vec{v}+\vec{u} = \vec{v}_{\textmd{out}} \\
(\lambda,0) & \textrm{ otherwise.}
\end{array} \right.
$$
Let $T_P = \left\{ \left. t_{\vec{v}} \; \right| \vec{v} \in P
\right\}$. Note that since $P$ is finite, so too is $T_P$. Now
define the TAS $\mathcal{T}_P = (T_P,\sigma_P,1)$, where for all
$\vec{v} \in \mathbb{N}^2$, $\sigma_P$ is defined as
$$
\sigma_P(\vec{v}) = \left\{
\begin{array}{ll}
t_{(0,0)} & \textrm{ if } \vec{v} = (0,0) \\
\uparrow & \textrm{ otherwise.}
\end{array} \right.
$$
It is routine to verify that $\mathcal{T}_P$ is directed (i.e., $P$
uniquely self-assembles in $\mathcal{T}_P$).
\end{construction}

We now have the machinery to prove our first impossibility result.

\begin{theorem}
\label{weaktheorem} If $F \subsetneq \mathbb{N}^2$ is a discrete
self-similar fractal, $\fgg{F}$ is connected, and $F$ fully weakly
self-assembles in the TAS $\mathcal{T}_F = (T,\sigma,\tau)$, where
$\sigma$ consists of a single tile placed at the origin, then $\tau
> 1$.
\end{theorem}
\begin{proof}
Suppose that $F$ is generated by the set $V \subseteq \{0,\ldots
c-1\}^2$, and assume for the sake of obtaining a contradiction that
$\tau = 1$. Let $V' = \{0,\ldots c-1\}^2-V$. There are two cases to
consider.
\begin{description}
\item[Case 1] If there exists a
path $P = \langle (x_0,y_0),\ldots, (x_{l-1},y_{l-1})\rangle$ in
$\fgg{V'}$, with $\fgg{P}$ connected, satisfying either of the
following.
\begin{enumerate}
\item $(x_0,y_0) \in (\{0\}\times\{0,\ldots c-1\})$ and
$(x_{l-1},y_{l-1}) \in (\{c-1\}\times\{0,\ldots c-1\})$.
\item $(x_0,y_0) \in (\{0\ldots,c-1\}\times\{0\})$ and
$(x_{l-1},y_{l-1}) \in (\{0,\ldots,c-1\}\times\{c-1\})$.
\end{enumerate}

Without loss of generality, assume that $P$ satisfies (1). First
note that there exists $\vec{a} \in V$, and there is no path from
$(0,0)$ to $\vec{a}$ in $\fgg{V}$. Define, for all $i \in
\mathbb{N}$, the points
$$
\vec{a}_i = c^i \cdot \vec{a}.
$$
Since $F$ is infinite, it is possible to choose $k \in \mathbb{N}$
large enough so that the path
$$
P = \left\langle (x_0,y_0),(x_1,y_1), \ldots
(x_{k-1},y_{k-1})\right\rangle
$$
satisfies the following properties.
\begin{enumerate}
\item $(x_0,y_0) = (0,0)$,
\item there exists $l \in \mathbb{N}$ such that $(x_{k-1},y_{k-1}) = \vec{a}_l$,
\item $\fgg{P}$ is connected and simple (in fact a tree), and
\item there exists a sub-path $P' \subset P$, such that $\fgg{P'}$ is
connected, $P' \subseteq \mathbb{N}^2 - F$, and $|P'| > 12|T|$
(because $F$ fully weakly self-assembles).
\end{enumerate}

Since $\tau = 1$, there is an assembly sequence $\vec{\alpha} =
(\alpha_i \mid 0\leq i < k)$, with $\alpha = \res{\vec{\alpha}}$,
satisfying $\alpha_0 = \sigma$, and $\dom{\alpha} = P$. Then by
Construction~\ref{crapfactory2} there exists a $1$-$T_P$-assembly
sequence $\vec{\alpha}_P = (\alpha_i \mid 0\leq i < k)$, with result
$\alpha_P = \res{\vec{\alpha}_P}$ satisfying $\dom{\alpha_P} = P$,
and $\alpha_P(x_{l-1},y_{l-1}) \in B$. By (4), there exist
$\vec{s},\vec{t} \in P'$ such that $\alpha_P(\vec{s}) =
\alpha_P(\vec{t})$, and $\vec{s},\vec{t} \not \in F$. Let
$P_{\dom{\sigma},\vec{s}}$ and $P_{\dom{\sigma},\vec{t}}$ be
$\dom{\sigma}$-subgraphs of $P$. Then Lemma~\ref{pathlemma} tells us
that $P_{\dom{\sigma},\vec{s}} \sim P_{\dom{\sigma},\vec{t}}$,
whence there exists a location $\vec{b} \in P'$ such that
$\alpha_P(\vec{b}) \in B$. This contradicts the definition of $P$.

\item[Case 2] If there is no such path in $\fgg{V'}$, then we proceed as follows.
First note that there exists $\vec{a} \not \in V$. It is clear that,
for all $i \in \mathbb{N}$, $c^i\cdot \vec{a}+(1,1) \not \in F$. For
each $i \in \mathbb{N}$, define the point
$$
\vec{a}_i = c^i \cdot \vec{a}+(1,1).
$$

Since $F$ is infinite, it is possible to choose $k \in \mathbb{N}$
large enough so that the path
$$
P = \left\langle (x_0,y_0),(x_1,y_1), \ldots
(x_{k-1},y_{k-1})\right\rangle
$$
satisfies the following properties.
\begin{enumerate}
\item $(x_0,y_0) = (0,0)$,
\item there exists $l \in \mathbb{N}$ such that $(x_{k-1},y_{k-1}) = \vec{a}_l$ (because $F$ fully weakly self-assembles),
\item $\fgg{P}$ is connected and simple (in fact a tree), and
\item for all $\vec{u} \in U_2$, $\min\left\{ i \mid i\cdot \vec{u} + \vec{a}_l \in F \right\} > 12|T|$.
\end{enumerate}
Since $\tau = 1$, there is an assembly sequence $\vec{\alpha} =
(\alpha_i \mid 0\leq i < k)$, with $\alpha = \res{\vec{\alpha}}$,
satisfying $\alpha_0 = \sigma$, and $\dom{\alpha} = P$. Then by
Construction~\ref{crapfactory2} there exists a $1$-$T_P$-assembly
sequence $\vec{\alpha}_P = (\alpha_i \mid 0\leq i < k)$, with result
$\alpha_P = \res{\vec{\alpha}_P}$ satisfying $\dom{\alpha_P} = P$.
By (4), there exist $\vec{s},\vec{t} \in P$ such that
$\alpha_P(\vec{s}) = \alpha_P(\vec{t})$, and $\vec{s},\vec{t} \not
\in F$. Let $P_{\dom{\sigma},\vec{s}}$ and
$P_{\dom{\sigma},\vec{t}}$ be $\dom{\sigma}$-subgraphs of $P$. Then
Lemma~\ref{pathlemma} tells us that $P$ can be extended to an
infinite, periodic path $P'$ consisting of all but finitely many
non-black tiles (i.e., tiles that are placed on the points in
$\mathbb{N}^2 - F$). This contradicts the definition of $F$.
\end{description}
\end{proof}

Note that Theorem~\ref{firstmaintheorem} says that even if one is
allowed to place a tile at {\it every} location in the first
quadrant, it is still impossible for self-similar fractals to weakly
self-assemble at temperature 1.

Next, we exhibit a class $\mathcal{C}$ of (non-tree) ``pinch-point''
discrete self-similar fractals that do not strictly self-assemble.
Before we do so, we establish the following lower bound.

\begin{lemma}
\label{lowerbound} If $X \subseteq \mathbb{Z}^2$ strictly
self-assembles in the TAS $\mathcal{T} = (T,\sigma,\tau)$, where
$\sigma$ consists of a single tile placed at the origin, then $|T|
\geq \left| \left\{ B \; \left| \; B \textmd{ is a unique }
\dom{\sigma}\textmd{-subgraph of } \fgg{X} \right. \right\}
\right|$.
\end{lemma}
\begin{proof}
Assume the hypothesis, and let $\alpha \in \termasm{\mathcal{T}}$.
For the purpose of obtaining a contradiction, suppose that $|T| <
\left| \left\{ B \; \left| \; B \textmd{ is a unique }
\dom{\sigma}\textmd{-subgraph of } \fgg{X} \right. \right\}
\right|$. By the Pigeonhole Principle, there exists points
$\vec{r},\vec{r}' \in X$ satisfying (1) $\alpha\left(\vec{r}\right)
= \alpha\left(\vec{r}'\right)$, and (2) $G_{\dom{\sigma},\vec{r}}
\not \sim G_{\dom{\sigma},\vec{r}'}$. Let $\sigma'$ be the assembly
with $\dom{\sigma'} = \left\{\vec{r}'\right\}$, and for all $\vec{u}
\in U_2$, define
$$
\sigma'(\vec{r}')(\vec{u}) = \left\{
\begin{array}{ll}
\left(\color_{\alpha(\vec{r}')}(\vec{u}),\strength_{\alpha(\vec{r}')}(\vec{u})\right) & \textrm{ if } \vec{r}'+\vec{u} \in G_{\dom{\sigma},\vec{r}'} \\
(\lambda,0) & \textrm{ otherwise.}
\end{array} \right.
$$
Let $\sigma''$ be the assembly with $\dom{\sigma''} = \left\{
\vec{r}'' \right\}$, and for all $\vec{u} \in U_2$, define
$$
\sigma'(\vec{r}'')(\vec{u}) = \left\{
\begin{array}{ll}
\left(\color_{\alpha(\vec{r}'')}(\vec{u}),\strength_{\alpha(\vec{r}'')}(\vec{u})\right) & \textrm{ if } \vec{r}''+\vec{u} \in G_{\dom{\sigma},\vec{r}''} \\
(\lambda,0) & \textrm{ otherwise.}
\end{array} \right.
$$
Then $\mathcal{T}' = (T,\sigma,\tau)$ is a TAS in which
$G_{\dom{\sigma},\vec{r}'}$ strictly self-assembles, and
$\mathcal{T}'' = (T,\sigma'',\tau)$ is a TAS in which
$G_{\dom{\sigma},\vec{r}''}$ strictly self-assembles. But this is
impossible because $\alpha\left(\vec{r}'\right) =
\alpha\left(\vec{r}''\right)$ implies that, for all $\vec{u} \in
U_2$, $\sigma'\left(\vec{r}'\right)\left(\vec{u}\right) =
\sigma''\left(\vec{r}''\right)\left(\vec{u}\right)$.
\end{proof}

Our lower bound is not as tight as possible, but it applies to a
general class of fractals. Our second impossibility result is the
following.

\begin{theorem}
\label{firstmaintheorem} If $X \subsetneq \mathbb{N}^2$ is a discrete
self-similar fractal satisfying (1) $\{(0,0),(0,c-1),(c-1,0)\}
\subseteq V$, (2) $V \cap (\{1,\ldots c-1\}\times \{c-1\}) =
\emptyset$, (3) $V \cap (\{c-1\} \times \{1,\ldots, c-1\}) =
\emptyset$, and (4) $\fgg{V}$ is connected, then $X$ does not
strictly self-assemble in the Tile Assembly Model.
\end{theorem}
\begin{proof}
By Lemma~\ref{lowerbound}, it suffices to show that, for any $m \in
\mathbb{N}$,
$$
\left|\left\{ B \; \left| \; B \textmd{ is a unique }
\dom{\sigma}\textmd{-subgraph of } \fgg{F} \right. \right\} \right|
\geq m.
$$
Define the points, for all $k \in \mathbb{N}$, $\vec{r}_k =
c^k(c(c-1),c-1)$, and let
$$
B_k = \left\{ (a,b) \in F \left| (a,b) \in \{0,\ldots c^k-1\}^2 +
\vec{r}_k \right.\right\}.
$$
Conditions (1), (2), and (3) tell us that $\fgg{B_k}$ is a
$\dom{\sigma}$-subgraph of $\fgg{F}$ (rooted at $\vec{r}_k$), and it
is routine to verify that, for all $k,k' \in \mathbb{N}$ such that
$k \ne k'$, $\fgg{B_k} \not \sim \fgg{B_{k'}}$. Thus, we have
\begin{eqnarray*}
m & = & \left| \left\{ \left. \fgg{B_k} \; \right| 0 \leq k < m \right\} \right| \\
  & \leq & \left| \left\{ B \; \left| \; B \textmd{ is a unique } \dom{\sigma} \textmd{-subgraph of } \fgg{F} \right.\right\} \right|. \\
\end{eqnarray*}
\end{proof}

\begin{corollary}[Lathrop, et. al. \cite{SSADST}]
\label{corollary0}The standard discrete Sierpinski triangle
$\mathbf{S}$ does not strictly self-assemble in the Tile Assembly
Model.
\end{corollary}

%% file: section4.tex
\section{Every Nice Self-Similar Fractal Has a Fibered Version}
\begin{figure}
\begin{center}
\includegraphics[width=5.5in]{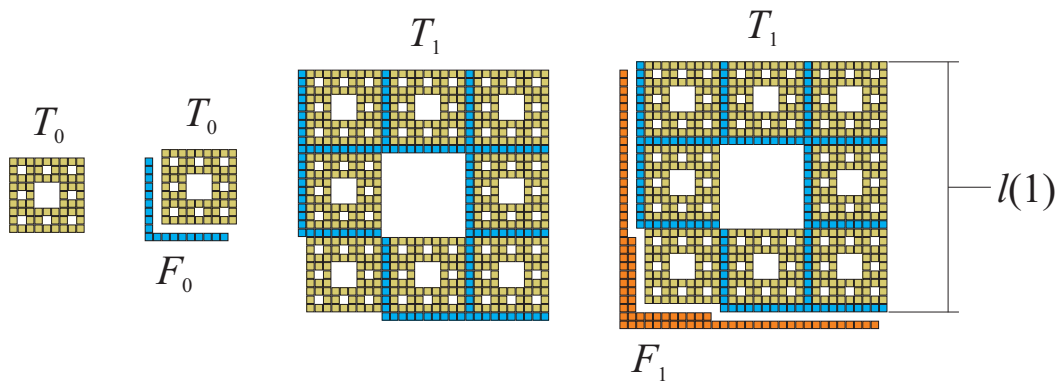}
  \caption{\small Construction of the fibered Sierpinski carpet. The blue, and orange tiles represent (possibly translated copies of) $F_0$, and $F_1$, respectively. Note that this image should be viewed in color.}
  \label{fig:fiberedexample}
  \vspace{-20pt}
\end{center}
\end{figure}

In this section, given a nice $c$-discrete self-similar fractal $X
\subsetneq \mathbb{N}^2$ (generated by $V$), we define its fibered
counterpart $\mathbf{X}$. Intuitively, $\mathbf{X}$ is nearly
identical to $X$, but each successive stage of $\mathbf{X}$ is
slightly thicker than the equivalent stage of $X$ (see
Figure~\ref{fig:fiberedexample} for an example). Our objective is to
define sets $F_0,F_1,\ldots \subseteq \mathbb{Z}^2$, sets
$T_0,T_1,\ldots \subseteq \mathbb{Z}^2$, and functions
$l,f,t:\mathbb{N} \rightarrow \mathbb{N}$ with the following
meanings.
\begin{enumerate}
\item $T_i$ is the $i^{\textmd{th}}$ stage of our construction of
the fibered version of $X$.
\item $F_i$ is the {\it fiber} associated with $T_i$. It is the
smallest set whose union with $T_i$ has a vertical left edge and a
horizontal bottom edge, together with one additional layer added to
these two now-straight edges.
\item $l(i)$ is the length (number of tiles in) the left (or bottom)
edge of $T_i \cup F_i$.
\item $f(i) = |F_i|$.
\item $t(i) = |T_i|$.
\end{enumerate}

These five entities are defined recursively by the equations
\begin{alignat*}{2}
&T_0 = X_2 \text{ (the third stage of $X$)}, \\
&F_0 = \left( \left\{ -1 \right\} \times  \left\{ -1, \ldots, c^2
\right\} \right) \cup \left( \left\{-1, \ldots, c^2 \right\} \times
\left\{-1\right\}
\right), \\
&l(0) = c^2+1, \; f(0) = 2c^2+1, \; t(0) = (|V|+1)^2, \\
&T_{i+1} = T_i \cup \left( \left( T_i \cup F_i \right) +l(i) V
\right),
\\
&F_{i+1} = F_i \cup \left( \left\{ -i-2 \right\} \times \left\{
-i-2, -i-1,
\cdots , l(i+1)-i-3\right\}\right) \\
&\cup \left( \left\{-i-2, -i-1, \cdots, l(i+1)-i-3 \right\} \times
\left\{-i-2
\right\} \right),  \\
&l(i+1) = c\cdot l(i) + 1, \\
&f(i+1) = f(i) + c\cdot l(i+1) - 1, \\
&t(i+1) = |V|t(i) + f(i).
\end{alignat*}
Finally, we let
$$
\displaystyle \mathbf{X} = \bigcup_{i=0}^{\infty}{T_i}.
$$

Note that the set $T_i \cup F_i$ is the union of an ``outer
framework,'' with an ``internal structure.'' One can view the outer
framework of $T_i \cup F_i$ as the union of a square $S_i$ (of size
$i+2$), a rectangle $X_i$ (of height $i+2$ and width $l(i)-(i+2)$),
and a rectangle $Y_i$ (of width $i+2$ and height $l(i)-(i+2)$).
Moreover, one can show that the internal structure of $T_i \cup F_i$
is simply the union of (appropriately-translated copies) of smaller
and smaller $X_i$ and $Y_i$-rectangles.

We have the following ``similarity'' between $X$ and $\mathbf{X}$.

\begin{lemma}
If $X \subsetneq \mathbb{N}^2$ is a nice self-similar fractal, then
$\textmd{Dim}_{\zeta}(X) = \textmd{Dim}_{\zeta}(\mathbf{X})$.
\end{lemma}

In the next section we sketch a proof that the fibered version of
every nice self-similar fractal strictly self-assembles.
\vspace{-10pt}

%% file: section5.tex
\section{Sketch of Main Construction}
Our second main theorem says that the fibered version of every nice
self-similar fractal strictly self-assembles in the Tile Assembly
Model (regardless of whether the latter strictly self-assembles).

\begin{theorem}
\label{secondmaintheorem} For every nice self-similar fractal $X
\subset \mathbb{N}^2$, there is a directed TAS in which $\mathbf{X}$
strictly self-assembles.
\end{theorem}

We now give a brief sketch of our construction of the singly-seeded
TAS $\mathcal{T}_{\mathbf{X}} = (X_{\mathbf{X}},\sigma,2)$ in which
$\mathbf{X}$ strictly self-assembles. The full construction is
implemented in C++, and is available at the following URL:
\url{http://www.cs.iastate.edu/~lnsa}.

Throughout our discussion, $S_{\vec{u}}$, $Y_{\vec{u}}$, and
$X_{\vec{u}}$ refer to the square, the vertical rectangle and the
horizontal rectangle, respectively, that form the ``outer
framework'' of the set $\left(\left(T_{i} \cup
F_{i}\right)+l(i)\cdot \vec{u}\right)$ (See the right-most image in
Figure~\ref{fig:algorithm}).

\subsection{Construction Phase 1}
\begin{figure}[htp]
\begin{center}
\includegraphics[width=5.0in]{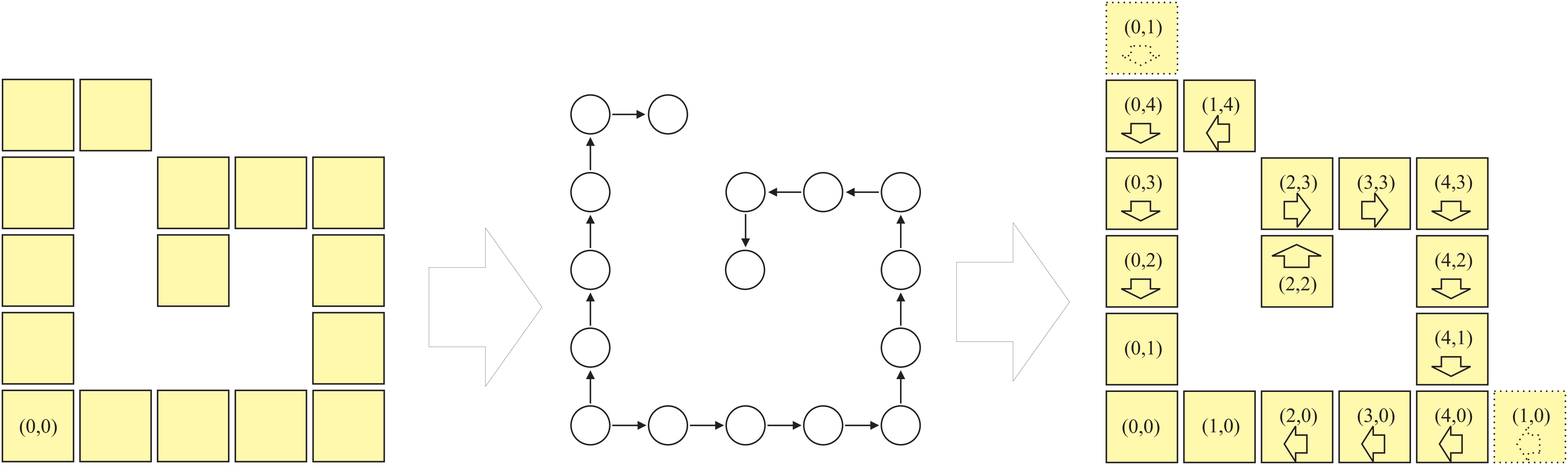}
\caption{\small Phase 1 of our construction. Notice the two special
cases (right-most image) in which we define $(0,1)_{\textmd{in}}$
and $(1,0)_{\textmd{in}}$.} \label{sp}
\end{center}
 \vspace{-10pt}
\end{figure}

Here, directed graphs are considered.  Let $X$ be a nice ($c$-discrete)
self-similar fractal generated by
$V$. We first compute a directed spanning tree $B = (V,E)$ of
$\fgg{V}$ using a breadth-first search, and then compute the graph
$B^{\textmd{R}} = \left(V,E^{\textmd{R}}\right)$, where
$$
E^{\textmd{R}} = \left\{ \left(\vec{v},\vec{u}\right) \mid
\left(\vec{u},\vec{v}\right) \in E \textmd{ and } \vec{u} \ne
(0,0)\right\} \cup \{ ((0,1),(0,c-1)),((1,0),(c-1),0)\}.
$$
Figure~\ref{sp} depicts phase 1 of our construction for a particular
nice self-similar fractal.

\begin{notation}
For all $\vec{0} \ne \vec{u} \in V$, $\vec{u}_{\textmd{in}}$ is the
unique location $\vec{v}$ satisfying $\left(\vec{u},\vec{v}\right)
\in E^{\textmd{R}}$.
\vspace{-10pt}
\end{notation}


\subsection{Construction Phase 2}
In the second phase we construct, for each $(0,0) \ne \vec{u} \in
V$, a finite set of tile types $T_{\vec{u}}$ that self-assemble a
particular subset of $\mathbf{X}$. There are two cases to consider.
\begin{description}
\item[Case 1] In the first case, we generate, for each $\vec{u} \in V -
\left\{(0,0),(0,1),(1,0)\right\}$, three sets of tile types
$T_{S_{\vec{u}}}$, $T_{X_{\vec{u}}}$, and $T_{Y_{\vec{u}}}$ that,
when combined together, and assuming the presence of
$\left(\left(T_i \cup F_i\right)+l(i)\cdot
\vec{u}_{\textmd{in}}\right)$, self-assemble the set
$\left(\left(T_i \cup F_i\right)+l(i)\cdot \vec{u}\right)$, for any
$i \in \mathbb{N}$.

\item[Case 2] In the second case, we generate, for each $\vec{u} \in
\{(0,1),(1,0)\}$, the same three sets of tile types
($T_{S_{\vec{u}}}$, $T_{X_{\vec{u}}}$, and $T_{Y_{\vec{u}}}$) that
self-assemble the set $\left(\left(T_{i} \cup F_{i}\right)+l(i)\cdot
\vec{u}\right)$ ``on top of'' the set $\left(\left(T_{i-1} \cup
F_{i-1}\right)+l(i-1)\cdot \vec{u}_{\textmd{in}}\right)$, for any $i
\in \mathbb{N}$.
\end{description}

\begin{figure}[htp]
\begin{center}
\psfrag{S}{$S_{\vec{u}}$} \psfrag{X}{$X_{\vec{u}}$}
\psfrag{Y}{$Y_{\vec{u}}$}
\includegraphics[width=5.0in]{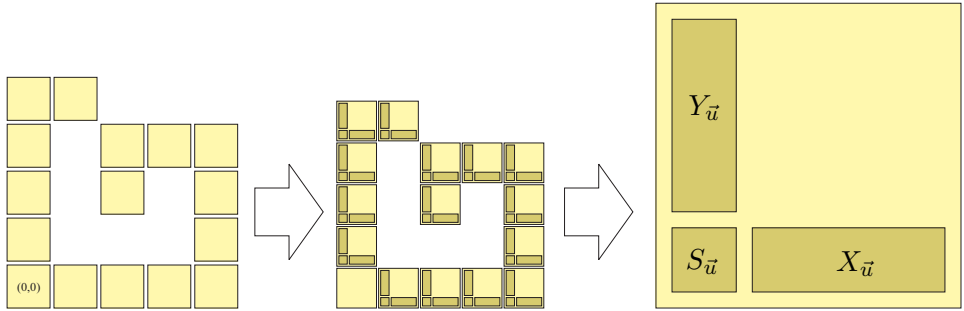}
\caption{\small Let $V$ be the left-most image. The first arrow
represents phase 2 of the construction. The second arrow shows a
magnified view of a particular point in $V$. Each point $(0,0) \ne
\vec{u} \in V$ can be viewed conceptually as three components: the
tile sets $T_{S_{\vec{u}}}$, $T_{X_{\vec{u}}}$ and $T_{Y_{\vec{u}}}$
that ultimately self-assemble the square $S_{\vec{u}}$, and the
horizontal and vertical rectangles $X_{\vec{u}}$ and $Y_{\vec{u}}$
respectively.} \label{fig:algorithm}
\end{center}
 \vspace{-20pt}
\end{figure}

Finally, we let $T_{\mathbf{X}} = \bigcup_{(0,0) \ne \vec{u} \in
V}{T_{\vec{u}}}$, where $T_{\vec{u}} = T_{S_{\vec{u}}} \cup
T_{X_{\vec{u}}} \cup T_{Y_{\vec{u}}}$. Figure~\ref{fig:algorithm}
gives a visual interpretation of the second phase of our
construction. Our TAS is $\mathcal{T}_{\mathbf{X}} =
(T_{\mathbf{X}},\sigma,2)$, where $\sigma$ consists of a single
``seed'' tile type placed at the origin. Our full construction
yields a tile set of 5983 tile types for the fractal generated by
the points in the left-most image in Figure~\ref{fig:algorithm}.

\subsection{Details of Construction}
Note that in our construction, the self-assembly of the
sub-structures $S_{\vec{u}}$, $Y_{\vec{u}}$, and $X_{\vec{u}}$ can
proceed either {\it forward} (away from the axes) or {\it backward}
(toward the axes).
\subsubsection{Forward Growth}
We now discuss the self-assembly of the set $((T_i \cup
F_i)+\vec{u}\cdot l(i))$ for $\vec{u} \in V$ satisfying
$\vec{u}_{\textmd{in}} \in \left( \vec{u}+ \{ (-1,0), (0,-1)\}
\right)$.
\begin{wrapfigure}{r}{0.25\textwidth}
\begin{center}
\includegraphics[width=.37in]{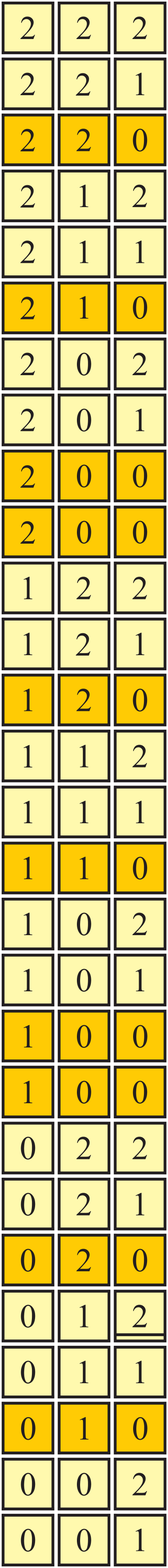} \label{wow}
\caption{\small Example of a base-3 modified binary counter. The
darker shaded rows are the spacing rows.}
\end{center}
\end{wrapfigure}
If $\vec{u} \not \in \{(0,0),(0,1),(1,0)\}$ (i.e., case 1 of phase
2), then the tile set $T_{S_{\vec{u}}}$ self-assembles the square
$S_{\vec{u}}$ directly on top (or to the right) of, and having the
same width (height) as, the rectangle $Y_{\vec{u}_{\textmd{in}}}$
($X_{\vec{u}_{\textmd{in}}}$). If $\vec{u} \in \{(0,1),(1,0)\}$
(i.e., case 2 of phase 2), then the tile set $T_{S_{\vec{u}}}$
self-assembles the square $S_{\vec{u}}$ on top (or to the right) of
the set $Y_{\vec{u}_\textmd{in}}$ such that right (top) edge of the
former is flush with that of the latter. Note that in case 2, the
width of $Y_{\vec{u}_{\textmd{in}}}$ is always one less than that of
$S_{\vec{u}}$. In either case, it is straightforward to construct
such a tile set $T_{S_{\vec{u}}}$.

The tile set of $T_{Y_{\vec{u}}}$ self-assembles a fixed-width
base-$c$ counter (based on the ``optimal'' binary counter presented
in \cite{CGM04}) that, assuming a width of $i \in \mathbb{N}$,
implements the following counting scheme: Count each positive
integer $j$, satisfying $1 \leq j \leq c^i-1$, in order but count
each number exactly
$$
\bval{c \textmd{ divides } j}\cdot \rho(j) + \bval{c \textmd{ does
not divide } j}\cdot 1
$$
times, where $\rho(j)$ is the largest number of consecutive
least-significant 0's in the base-$c$ representation of $j$, and
$\bval{\phi}$ is the {\it Boolean} value of the statement $\phi$.
The {\it value} of a row is the number that it represents. We refer
to any row whose value is a multiple of $c$ as a {\it spacing row}.
All other rows are {\it count} rows. The {\it type} of the counter
that self-assembles $Y_{\vec{u}}$ is $\vec{u}$.

Each counter self-assembles on top (or to the right) of the square
$S_{\vec{u}}$, with the width of the counter being determined by
that of the square. It is easy to verify that if the width of
$S_{\vec{u}}$ is $i+2$, then $T_{\vec{Y}_{\vec{u}}}$ self-assembles
a rectangle having a width of $i+2$ and a height of
$$
\left(c^2+1\right)c^i+\frac{c^i-1}{c-1} = l(i)-(i+2),
$$
which is exactly $Y_{\vec{u}}$. Figure~5 shows the counting scheme
of a base-$3$ counter of width $3$. We construct the set
$T_{X_{\vec{u}}}$ by simply reflecting the tile types in
$T_{Y_{\vec{u}}}$ about the line $y=x$, whence the three sets of
tile types $T_{S_{\vec{u}}}$, $T_{X_{\vec{u}}}$, and
$T_{Y_{\vec{u}}}$ self-assemble the ``outer framework'' of the set
$((T_i \cup F_i)+\vec{u}\cdot l(i))$.

The ``internal structure'' of the set $((T_i \cup F_i)+\vec{u}\cdot
l(i))$ self-assembles as follows. Oppositely oriented counters
attach to the right side of each contiguous group of spacing rows in
the counter (of type $\vec{u}$) that self-assembles $Y_{\vec{u}}$.
The number of such spacing rows determines the height of the
horizontal counter, and its type is $\left(0,j/c \mod c\right)$,
where $j$ is the value of the spacing rows to which it attaches. We
also hard code the glues along the right side of each non-spacing
row to self-assemble the internal structure of the points in the set
$T_0$.

The situation for $X_{\vec{u}}$ is similar (i.e., a reflection of
its vertical counterpart), with the exception that the glues along
the top of each non-spacing row are configured differently than they
were for $Y_{\vec{u}}$. This is because nice self-similar fractals
need not be symmetric.

One can prove that, by recursively attaching smaller
oppositely-oriented counters (of the appropriate type) to larger
counters in the above manner, the internal structure of $((T_i \cup
F_i)+\vec{u}\cdot l(i))$ self-assembles.

\subsubsection{Reverse Growth}
We now discuss the self-assembly of the set $((T_i \cup
F_i)+\vec{u}\cdot l(i))$, for all $\vec{u} \in V$ satisfying
$\vec{u}_{\textmd{in}} \in \left( \vec{u}+ \{ (1,0), (0,1)\}
\right)$.

\begin{figure}[htp]
\psfrag{Sa}{$S_{\vec{u}}$} \psfrag{X}{$X_{\vec{u}}$}
\psfrag{Xa}{$X_{\vec{u}}$} \psfrag{Sb}{$S_{\vec{u}_{\textmd{in}}}$}
\psfrag{Xb}{$X_{\vec{u}}$} \psfrag{Sc}{$S_{\vec{u}_{\textmd{in}}}$}
\psfrag{Xc}{$X_{\vec{u}}$}
  \begin{center}
  \subfloat[\;]{\label{fig:r1}\includegraphics[width=1.39in]{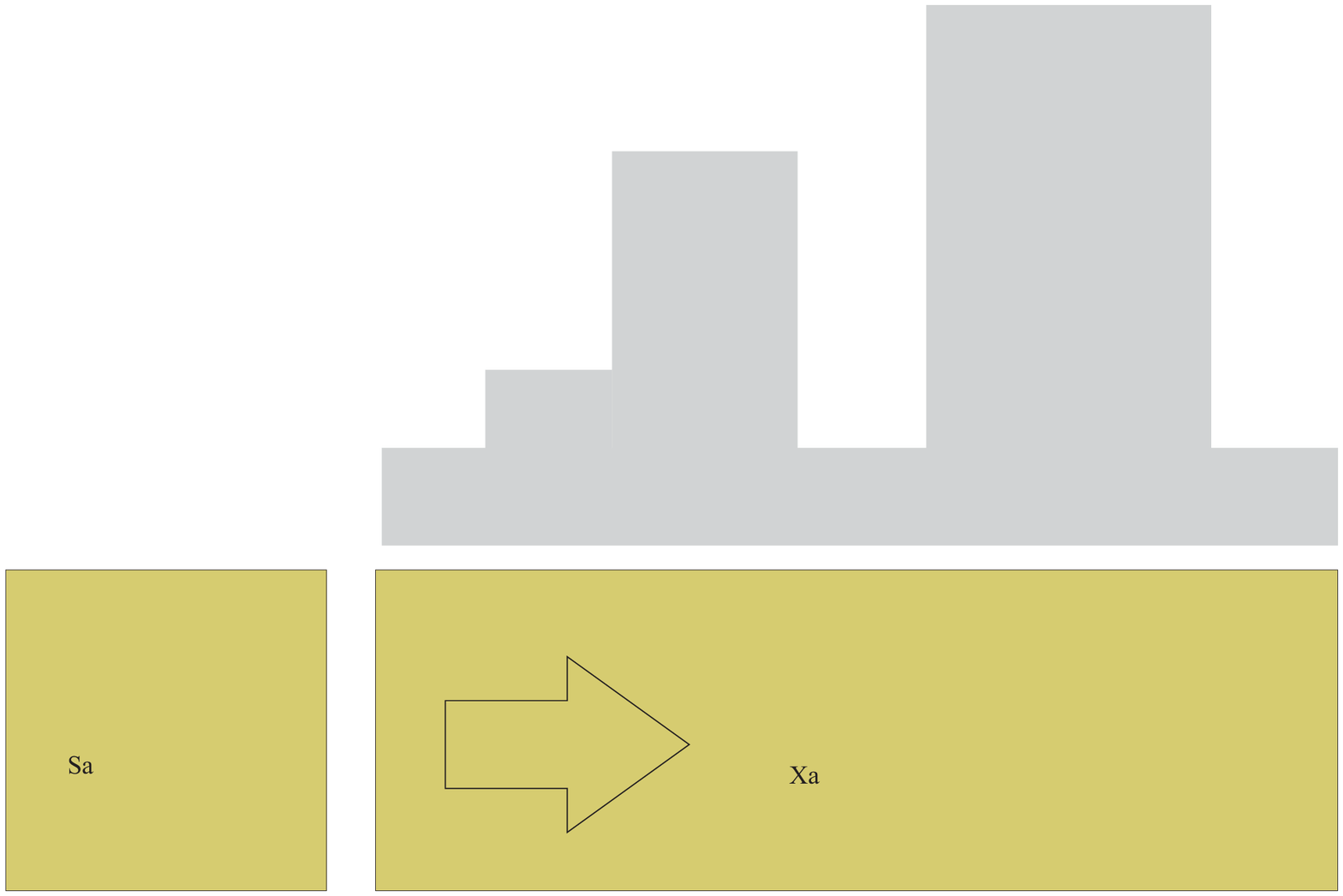}}
  \quad\quad
  \subfloat[\;]{\label{fig:r2}\includegraphics[width=1.39in]{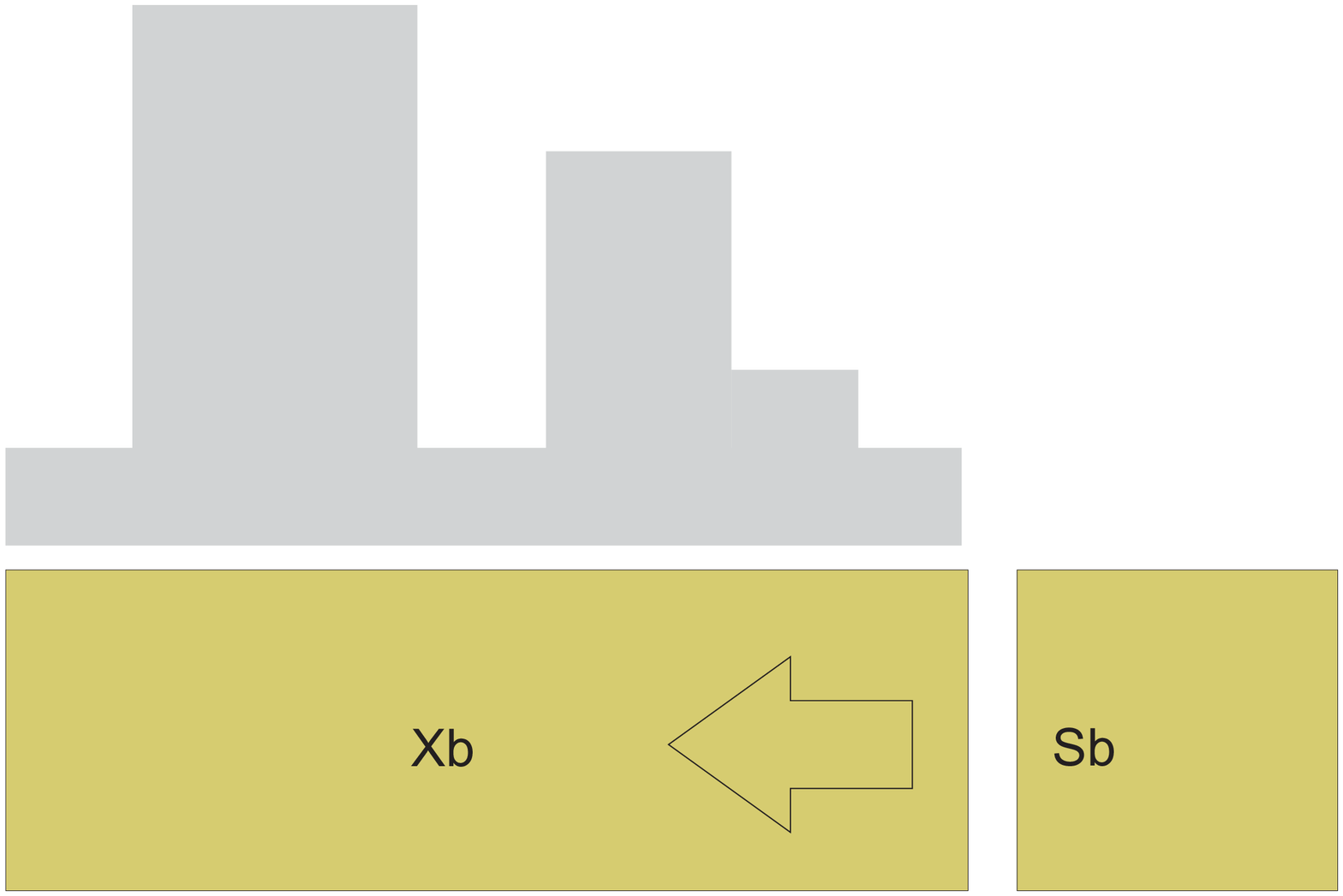}}
  \quad\quad
  \subfloat[\;]{\label{fig:r3}\includegraphics[width=1.39in]{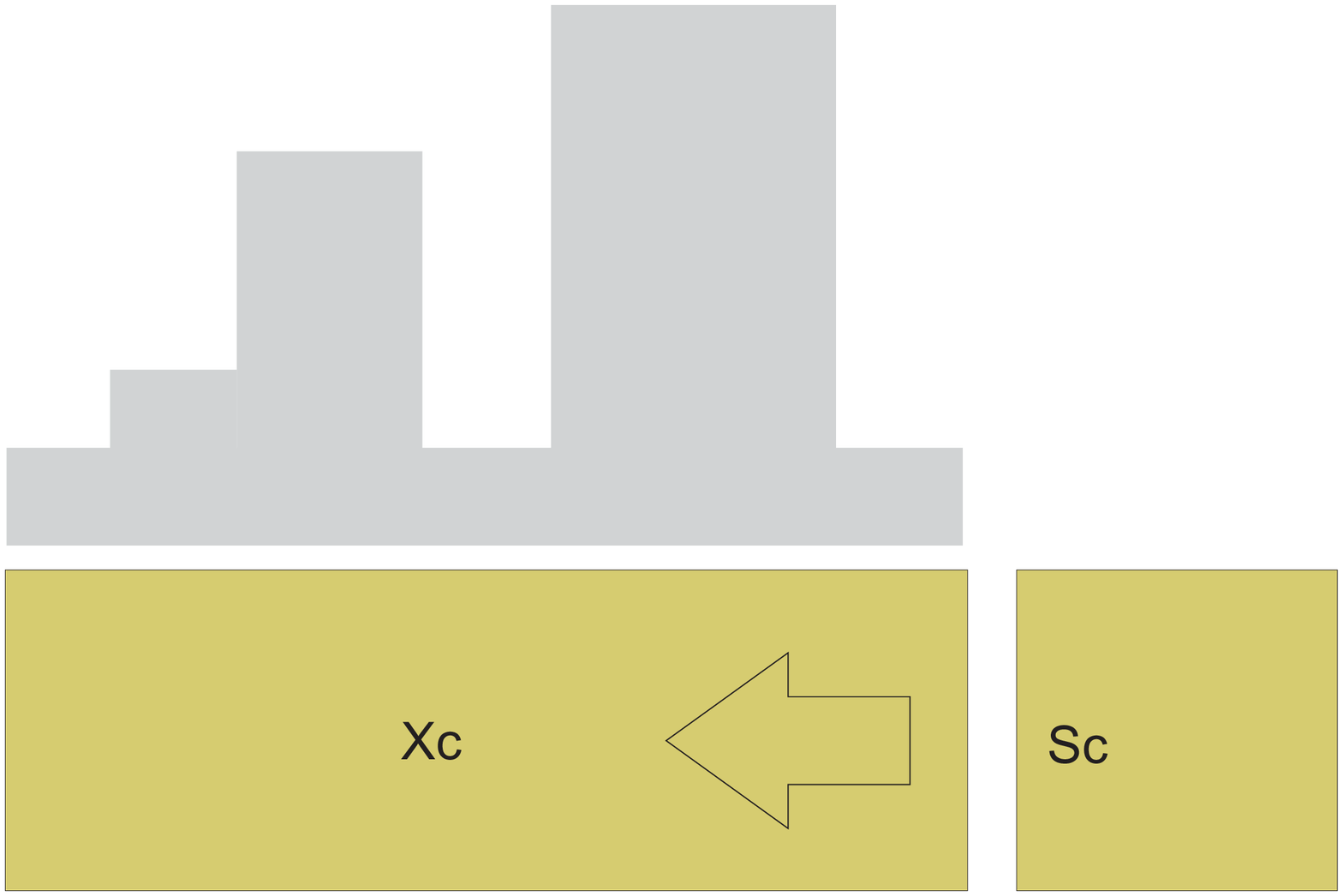}}
  \caption{\small (a) depicts forward growth, (b) shows what happens if the tile set $T_{X_{\vec{u}}}$ were to simply ``count in reverse,'' and (c) is the desired result. }
  \label{fig:reversegrowth}
 \vspace{-20pt}
  \end{center}
\end{figure}

In this case, the tile set $T_{Y_{\vec{u}}}$ ($T_{X_{\vec{u}}}$)
self-assembles the set $Y_{\vec{u}}$ ($X_{\vec{u}}$) directly below
(or to the left of) the square $S_{\vec{u}_{\textmd{in}}}$, and
grows toward the $x$-axis (or $y$-axis) according to the base-$c$
counting scheme outlined above. We also configure $T_{Y_{\vec{u}}}$
($T_{X_{\vec{u}}}$) so that the right (or top)-most edge of
$Y_{\vec{u}}$ ($X_{\vec{u}}$) is essentially the ``mirror'' image of
its forward growing counterpart (See
Figure~\ref{fig:reversegrowth}). This last step ensures that the
internal structure of $((T_i \cup F_i)+\vec{u}\cdot l(i))$
self-assembles correctly. Next, the square $S_{\vec{u}}$ attaches to
the bottom (or left)-most edge of $Y_{\vec{u}}$ ($X_{\vec{u}}$).
Finally, the set $X_{\vec{u}}$ ($Y_{\vec{u}}$) self-assembles via
forward growth from the left (or top) of the square $S_{\vec{u}}$.

\subsubsection{Proof of Correctness} To prove the correctness of our
construction, we use a local determinism argument. The details of
the proof are technical, and therefore omitted from this version of
the paper.

%% file: section6.tex
\section{Conclusion}
In this paper, we (1) established two new absolute limitations of
the TAM, and (2) showed that fibered versions of ``nice''
self-similar fractals strictly self-assemble. Our impossibility
results motivate the following question: Is there a discrete
self-similar fractal $X \subsetneq \mathbb{N}^2$ that strictly
self-assembles in the TAM? Moreover, our positive result leads us to
ask: If $X \subsetneq \mathbb{N}^2$ is a discrete self-similar
fractal, then is it always the case that $X$ has a ``fibered''
version $\mathbf{X}$ that strictly self-assembles, and that is
similar to $X$ in some reasonable sense?